\documentclass[preprintnumbers,superscriptaddress,amsmath,amssymb,prd,preprint,nofootinbib,aps]{revtex4-1}
\usepackage{graphicx}
\usepackage{amsfonts}
\usepackage[colorlinks,linkcolor=red,anchorcolor=blue,citecolor=green]{hyperref}
\usepackage{subfigure}
\usepackage{float}
\usepackage{enumerate}
\usepackage{amsmath}
\usepackage{amssymb}
\usepackage{mathrsfs}
\usepackage{amsthm}

\begin{document}
\title{Superradiant instability of the Kerr-like black hole in Einstein-bumblebee gravity}
\author{Rui Jiang}
\author{Rui-Hui Lin}
\email[]{linrh@shnu.edu.cn}
\author{Xiang-Hua Zhai}
\email[]{zhaixh@shnu.edu.cn}
\affiliation{Division of Mathematics and Theoretical Physics, Shanghai Normal University, 100 Guilin Road, Shanghai 200234, China}

\begin{abstract}
An exact Kerr-like solution has been obtained recently in Einstein-bumblebee gravity model where Lorentz symmetry is spontaneously broken. In this paper, we investigate the superradiant instability of the Kerr-like black hole under the perturbation of a massive scalar field. We find the Lorentz breaking parameter $L$ does not affect the superradiance regime or the regime of the bound states. However, since $L$ appears in the metric and its effect cannot be erased by redefining the rotation parameter $\tilde{a}=\sqrt{1+L}a$, it indeed affects the bound state spectrum and the superradiance. We calculate the bound state spectrum via the continued-fraction method and show the influence of $L$ on the maximum binding energy and the damping rate. The superradiant instability could occur since the superradiance condition and the bound state condition could be both satisfied. Compared with Kerr black hole, the nature of the superradiant instability of this black hole depends non-monotonously not only on the rotation parameter of the black hole $\tilde{a}$ and the product of the black hole mass $M$ and the field mass $\mu$, but also on the Lorentz breaking parameter $L$. Through the Monte Carlo method, we find that for $l=m=1$ state the most unstable mode occurs at $L=-0.79637$, $\tilde{a}/M=0.99884$ and $M\mu=0.43920$, with the maximum growth rate of the field $\omega_{I}M=1.676\times10^{-6}$, which is about 10 times of that in Kerr black hole.

\end{abstract}
\maketitle

\section{Introduction}
\label{intro}
Lorentz invariance is one of the most important symmetries in General Relativity (GR) that works well in describing gravitation at the classical level. However, Lorentz invariance may not be an exact symmetry at all energy scales\cite{Mattingly_2005}. It is shown that Lorentz invariance will be strongly violated at the Planck scale ($\sim 10^{19}$ GeV) in some theories of quantum gravity(QG)\cite{Amelino_Camelia_2013}. Therefore, Lorentz violation in gravitation theories is worth studying in anticipation of a deeper understanding of nature. And hence, various theories involving Lorentz violation, such as ghost condensation\cite{Hamed_2004}, warped brane world, and Einstein-aether theory\cite{Jacobson_2001,Eling:2004dk,Jacobson:2007veq}, have been proposed and investigated. Moreover, the possibility of spontaneous Lorentz symmetry breaking(LSB) was considered. In 1989, Kosteleck{\'{y}} and Samuel presented a potential mechanism for the Lorentz breaking that may be generic in many string theories\cite{Kostelecky:1988zi,Kostelecky:1991ak}. The main idea is to find a model containing the essential features of the effective action that would arise in a string theory with tensor-induced breaking. One of the simplest ways to implement this idea is the so-called Einstein-bumblebee gravity. In this theory, a vector field ruled by a potential acquires a non-zero vacuum expectation value (VEV). The vector field is then frozen at its VEV, which chooses a preferred spacetime direction in the local frames and spontaneously breaks the Lorentz symmetry.

Since bumblebee gravity can be viewed as an endeavor to explore QG, it is then important to search for black hole solutions in this theory in that the strong gravitation environment around a black hole may provide information about QG. Casana \emph{et al} obtained an exact Schwarzschild-like solution in 2018\cite{Casana_2018}. Then, the light deflection\cite{PhysRevD.101.124058,_vg_n_2018,Carvalho:2021jlp} and quasinormal modes\cite{Oliveira:2021abg} of this black hole have been addressed. Moreover, spherically symmetric black hole solutions with cosmological constant\cite{PhysRevD.103.044002}, global monopole\cite{Gullu:2020qzu}, or Einstein-Gauss-Bonnet term\cite{Ding:2021iwv} have also been found. As a more practical scenario, axial symmetry in Einstein-bumblebee gravity is investigated and an exact Kerr-like black hole is obtained by Ding \emph{et al}\cite{Ding_2020}. The studies related to this solution have been extended to the effects of the matter and light around it\cite{Liu:2019mls,Wang:2021irh,Kanzi:2021cbg}. Furthermore, a Kerr-Sen-like black hole has also been found\cite{Jha:2020pvk}.

On the other hand, the study of black hole stability dates back to 1957. Regge and Wheeler proved that the Schwarzschild black hole is stable under small perturbations of the metric\cite{Regge:1957td,Vishveshwara:1970cc}. In the following studies, the propagation of the Klein-Gordon field around a black hole becomes an important tool for investigating the stability of the corresponding black hole. For a rotating black hole, the perturbing bosons may extract rotational energy from the black hole through the superradiance mechanism\cite{Starobinsky:1973aij,Bardeen:1972fi,Teukolsky:1974yv}. For this to occur, it is required that the frequency $\omega$ of the wave is smaller than a critical value determined by the azimuthal number $m$ of the perturbation and the angular velocity $\Omega_{H}$ of the black hole horizon
\begin{equation}\label{1}
	\omega<\omega_{c}\equiv m\Omega_{H}.
\end{equation}
If this superradiant condition is satisfied, the scattered wave will be amplified, where the additional energy comes from the black hole. Furthermore, if the superradiant process occurs repeatedly, the rotational energy of the black hole may be extracted for multiple times, triggering the superradiant instability of the black hole. This can be fulfilled with the ``black hole bomb" idea postulated by Press and Teukolsky\cite{Press:1972zz}, by arranging a special mirror surrounding the black hole to reflect the wave between the mirror and the black hole back and forth. Superradiant instability triggered by a mirror-like boundary condition has been studied extensively\cite{Cardoso:2004nk,Herdeiro:2013pia,Dolan:2015dha,Dias:2018zjg,Hod:2016rqd,Li:2020vid}. Besides these arranged walls to reflect the scattered waves, a natural wall may exist if the spacetime is asymptotically non-flat. For example, an anti-de Sitter spacetime is able to provide such reflection and help trigger the superradiant instability\cite{Khodadi:2020cht,Zhang:2014kna,Rahmani:2020vvv,Cardoso:2006wa,Li:2012rx,Zhu:2014sya,Destounis:2019hca,Huang:2016zoz}. Apart from being reflected by walls, scattered waves may be pulled back to the black hole if the perturbing boson has nonzero rest mass. The massive waves can naturally form bound states in the gravitational system and hence may trigger the superradiant instability\cite{Brito2015,Mehta:2021pwf,Witek:2012tr,Huang:2018qdl,Vieira:2021nha,Huang:2016qnk,Huang:2017nho}.

In this paper, we would like to investigate the stability of the Kerr-like black hole in the bumblebee gravity. We intend to study the bound states of the massive scalar field in the background of the Kerr-like black hole and hence the superradiant instability of the black hole. We obtain the frequency regime of the bound states and compute the bound state spectrum. When superradiant instability occurs, the scalar field will grow over time. In this case, there may be such a specific frequency of the bound state that this process will progress most rapidly. For Kerr spacetime, the maximum growth rate of the scalar field was found to be $\omega_{I}M=1.72\times10^{-7}$\cite{Dolan_2013}. On this basis, we will look into the influence of the LSB parameter $L$ on the maximum growth rate.

The paper is organized as follows. In Sec.\uppercase\expandafter{\romannumeral2}, we briefly review the Kerr-like black hole solution in
the bumblebee gravity. Then, we give the necessary conditions for the bound state in Sec.\uppercase\expandafter{\romannumeral3}. The bound state spectrum is computed by using the continued-fraction method in Sec.\uppercase\expandafter{\romannumeral4}. In Sec.\uppercase\expandafter{\romannumeral5}, we discuss the superradiant instability and further study the influence of the LSB parameter $L$ on the maximum growth rate. The last section is devoted to conclusion and discussions.

Throughout the paper, we follow the metric convention $\left(-,+,+,+\right)$  and the units $G=\hbar=c=1$.
\section{The Kerr-like black hole solution in the bumblebee gravity}
\label{Sec.2}
Under the framework of the bumblebee gravity theory, the spontaneous LSB is induced by a vector $B_{\mu}$ that has a non-zero VEV. The dynamics of the gravitational field will also be affected if it couples to $B_{\mu}$. Therefore, the action describing the LSB gravitation includes the free parts of both the gravitational field and the bumblebee field, the potential of $B_{\mu}$ that leads to a non-zero VEV, and the coupling between $B_{\mu}$ and the geometry of the spacetime. Based on the Riemannian description of the spacetime manifold, such an action can be written as\cite{Kosteleck__2004}
\begin{equation}
	\begin{aligned}
		S=\int d^{4} \sqrt{-g} \big [
		\frac{1}{16\pi } \left ( R+\xi B^{\mu }B^{\nu } R_{\mu \nu }   \right ) -\frac{1}{4} B^{\mu \nu }B_{\mu \nu }-V\left ( B_{\mu }B^{\mu}\pm b^2 \right ) \big ],
	\end{aligned}
\end{equation}
where $\xi$ is the coupling constant and the bumblebee field strength $B_{\mu\nu}$ is defined as
\begin{equation}
	B_{\mu \nu }=\partial _{\mu } B_{\nu}-\partial _{\nu} B_{\mu}.
\end{equation}
The potential $V$ takes a minimum at $B_{\mu }B^{\mu}\pm b^2=0$ with a real positive $b^2$, and therefore triggers the LSB of the bumblebee field. With $V=0$ and $V'=0$, the field $B_{\mu }$ is considered to be frozen at its VEV, i.e. $\left \langle B_{\mu}  \right \rangle =b_{\mu } $, where $b_{\mu }b^{\mu}=b^2$. And the field equation is then
\begin{equation}
	\begin{aligned}
		0=&R_{\mu \nu}-\kappa b_{\mu \alpha }b^{\alpha}_{\ \nu }+\frac{\kappa }{4} g_{\mu \nu } b^{\alpha \beta }b_{\alpha \beta } +\xi b_{\mu } b^{\alpha }R_{\alpha \nu } \\
		&+\xi b_{\nu } b^{\alpha }R_{\alpha \mu }-\frac{\xi }{2} g_{\mu \nu } b^{\alpha }b^{\beta}R_{\alpha\beta}-\frac{\xi }{2} \nabla _{\alpha }\nabla _{\mu} \left ( b^{\alpha }b_{\nu } \right ) \\
		&-\frac{\xi }{2} \nabla _{\alpha }\nabla _{\nu} \left ( b^{\alpha }b_{\mu } \right )+\frac{\xi }{2} \nabla^{2}\left ( b_{\mu } b_{\nu }  \right ) ,
	\end{aligned}
\end{equation}
where $\kappa=8\pi$.

For the axially symmetric scenario, by assuming the bumblebee field to be a purely radial form of $b_{\mu } =(0,b(r,\theta ),0,0)$, a Kerr-like black hole solution is found in Ref.\cite{Ding_2020}. In the standard Boyer-Lindquist coordinates, this solution is written as
\begin{equation}\label{metric}
	\begin{aligned}
	ds^{2}=&-\left(1-\dfrac{2Mr}{\rho^{2}} \right)dt^{2}+\dfrac{\rho^{2}}{\Delta}dr^{2}+\rho^{2}d\theta^{2}\\&+\dfrac{A\sin^{2}\theta}{\rho^{2}}d\varphi^{2}-\dfrac{4Mra\sqrt{1+L}\sin^{2}\theta}{\rho^{2}}dtd\varphi,
\end{aligned}
\end{equation}
where
\begin{equation}\label{metric1}
	\begin{aligned}
	&\rho^{2}=r^{2}+\left(1+L \right)a^{2}\cos^{2}\theta,\\
	&\Delta=\dfrac{r^{2}-2Mr}{1+L}+a^{2},\\
	&A=\left[r^{2}+\left(1+L \right)a^{2}  \right]^{2}-\Delta\left( 1+L\right)^{2}a^{2}\sin^{2}\theta.
\end{aligned}
\end{equation}
Here $L=\xi b^2$, $M$ is the ADM mass and $a$ is the Boyer-Lindquist parameter. Eqs.\eqref{metric} and\eqref{metric1} can be written in a form very resemblant to the Kerr solution
\begin{equation}\label{metric2}
	\begin{aligned}
		ds^{2}=&-\left(1-\dfrac{2Mr}{\tilde{\rho}^{2}} \right)dt^{2}+\dfrac{\tilde{\rho}^{2}}{\tilde{\Delta}}dr^{2}+\tilde{\rho}^{2}d\theta^{2}\\&+\dfrac{\tilde{A}\sin^{2}\theta}{\tilde{\rho}^{2}}d\varphi^{2}-\dfrac{4Mr\tilde{a}\sin^{2}\theta}{\tilde{\rho}^{2}}dtd\varphi,
	\end{aligned}
\end{equation}
where
\begin{equation}\label{metric3}
	\begin{aligned}
		&\tilde{a}=\sqrt{1+L}a,\\
		&\tilde{\rho}^{2}=r^{2}+\tilde{a}^{2}\cos^{2}\theta,\\
		&\tilde{\Delta}=\dfrac{r^{2}-2Mr+\tilde{a}^{2}}{1+L},\\
		&\tilde{A}=\left(r^{2}+\tilde{a}^{2}  \right)^{2}-(r^{2}-2Mr+\tilde{a}^{2})\tilde{a}^{2}\sin^{2}\theta.
	\end{aligned}
\end{equation}

Note that Eq.\eqref{metric2} now looks exactly the same as the Kerr metric in form. The cost, however, is that $\tilde{\Delta}$ still involves explicitly the LSB parameter $L$ that cannot be absorbed in the new spin parameter $\tilde{a}$. So, the effect of $L$ via $\tilde{\Delta}$ cannot be erased by redefining $\tilde{a}$. Hence the metric written in the form of Eqs.\eqref{metric2} and \eqref{metric3} is essentially different from the Kerr metric as long as the purely radial LSB parameter $L$ is nonzero. The event horizons locate at
\begin{equation}\label{9}
	\begin{aligned}
	&r_{\pm}=M\pm\sqrt{M^{2}-(1+L)a^{2}}=M\pm\sqrt{M^{2}-\tilde{a}^{2}},
\end{aligned}
\end{equation}
where $r_{+}$ and $r_{-}$ correspond to outer and inner horizons, respectively. The condition for the black hole to exist is then given by
\begin{equation}\label{amax}
	\sqrt{1+L}a= \tilde{a}  \le M .
\end{equation}
And the angular velocity of the outer horizon is
\begin{equation}\label{11}
	\Omega_{H}=\dfrac{a\sqrt{1+L}}{r_{+}^2 +(1+L)a^2}=\dfrac{\tilde{a}}{r_{+}^2 +\tilde{a}^2}.
\end{equation}
Since $\tilde{a}$ plays a similar role to the Boyer-Lindquist parameter $a$ in the Kerr metric, these physical quantities \eqref{9}-\eqref{11} and the superradiant condition\eqref{1} also have the same forms via $\tilde{a}$ as the Kerr black hole, and do not rely on $a$ or $L$ separately. In this sense, the effect of the LSB parameter $L$ via these measurable quantities is not trackable. However, as is seen in Eq.\eqref{metric3}, the effect of $L$ on the metric via $\tilde{\Delta}$ cannot be erased. Therefore, we will proceed with $\tilde{a}$ instead of the original Boyer-Lindquist parameter $a$, and will see in the following sections that the LSB parameter $L$ indeed affects the superradiance.

\section{Frequency regime of the bound states}
We consider a scalar field $\Psi$ with mass $\mu$ propagating in the background of the Kerr-like black hole \eqref{metric}, which is described by the Klein-Gordon equation
\begin{equation}\label{K-G}
	\dfrac{1}{\sqrt{-g}}\partial _{\mu }\left(g^{\mu\nu}\sqrt{-g}\partial_{\nu}\Psi \right)=\mu^2 \Psi.
\end{equation}
With the separation of $\Psi$,
\begin{equation}
	\Psi=R_{lm}(r)S_{lm}(\theta)e^{im\phi}e^{-i\omega t},
\end{equation}
Eq.(\ref{K-G}) can be decomposed into a radial part
\begin{equation}\label{radial}
	\tilde{\Delta}\frac{\mathrm{d} }{\mathrm{d} r} \left ( \tilde{\Delta}\frac{\mathrm{d}R_{lm} }{\mathrm{d} r} \right )  +UR_{lm}=0
\end{equation}
with
\begin{equation}
	\begin{aligned}\label{U}
		U=\frac{[\omega(r^{2} +\tilde{a}^{2}) -\tilde{a}m] ^{2}}{1+L}+\tilde{\Delta} (2m\tilde{a}\omega -A_{lm}-\tilde{a}^{2}\omega ^{2}-\mu ^{2}r^{2} ) ,
	\end{aligned}
\end{equation}
and an angular part
\begin{equation}\label{angular}
	\begin{aligned}
		\frac{1}{\sin \theta } \frac{\mathrm{d} }{\mathrm{d} \theta } \left ( \sin \theta \frac{\mathrm{d}S_{lm} }{\mathrm{d} \theta }\right ) +\bigg[ A_{lm}+\tilde{a}^{2}\left ( \omega ^{2} -\mu ^{2}\right ) \cos ^{2} \theta -\frac{m^{2} }{\sin ^{2} \theta } \bigg]S_{lm}=0,
	\end{aligned}
\end{equation}
where $R_{lm}$'s are radial functions, $S_{lm}$'s are spheroidal harmonic functions\cite{50831}, and $A_{lm}$'s are the angular eigenvalues that may be expanded as a power series
\begin{equation}
	A_{lm} =l(l+1)+\sum_{k=1}^{\infty } f_{k}\left [ \tilde{a}^{2}\left ( \mu ^{2}-\omega ^{2} \right ) \right ]^{k}
\end{equation}
if $\tilde{a}^{2}\left ( \mu ^{2}-\omega ^{2} \right )\le m^2 $. In addition, eigenvalue $A_{lm}$ can also be computed numerically with continued-fraction method\cite{1985}.

The radial equation(\ref{radial}) can be further written in the form
\begin{equation}\label{radial2}
	\frac{\mathrm{d}^{2} \widetilde R}{\mathrm{d} x^2} +\widetilde{U}\widetilde{R}=0 ,
\end{equation}
where
\begin{equation}
	\widetilde{R}=\sqrt{r^2 +\tilde{a}^2}R_{lm},
\end{equation}
 and the tortoise coordinate $x$ is defined by
\begin{equation}
	\dfrac{\mathrm{d}x}{\mathrm{d}r}=\dfrac{r^2 +\tilde{a}^2}{\tilde{\Delta}(1+L)}.
\end{equation}
The effective potential $\widetilde{U}$ in Eq.(\ref{radial2}) reads
\begin{equation}\label{effect potential}
	\begin{aligned}
		\widetilde{U}=\dfrac{U(1+L)^2}{\left(r^2 +\tilde{a}^2 \right) ^2}-\dfrac{r^2 \tilde{\Delta}^2 (1+L)^2}{\left(r^2 +\tilde{a}^2 \right)^4}-\dfrac{\tilde{\Delta}(1+L)}{r^2 +\tilde{a}^2  }\left[ \dfrac{r\tilde{\Delta}(1+L)}{\left(r^2 +\tilde{a}^2 \right) ^2} \right] ' ,
	\end{aligned}
\end{equation}
where the prime represents the derivative with respect to $r$.
$\widetilde{U}$ tends to $\left ( \omega^2 -\mu^2  \right )  \left ( 1+L \right )$ at the spatial infinity and $\left ( \omega -m\Omega _{H}  \right )^{2}   \left ( 1+L \right )$ at the outer horizon. Therefore, Eq.(\ref{radial2}) can be treated asymptotically as a wave equation at the horizon or the spatial infinity. Since we are interested in the bound state solutions, the wave towards the spatial infinity should decay exponentially and the wave at the horizon should be purely ingoing. Hence, the asymptotic states of the wave are
\begin{equation}\label{boundary condition}
	R_{lm}\sim \left\{
	\begin{aligned}
		&e^{-i\left ( \omega -\omega _{c}  \right )\sqrt{1+L} x }\quad&,\quad x\to -\infty ,\\
		\dfrac{1}{x}&e^{-\sqrt{(\mu ^{2}-\omega ^{2})(1+L)} x} \quad&,\quad x\to +\infty,
	\end{aligned}
	\right.	
\end{equation}
and the frequency $\omega$ must satisfy
\begin{equation}
	\omega^2 <\mu^2.
\end{equation}
Besides, a trapping potential well outside the black hole is also required for bound states\cite{Hod_2012}. We then proceed to consider the existence of the trapping potential well. With a new radial function $\psi$ defined by
\begin{equation}
	\psi\equiv\tilde{\Delta}^{1/2}R_{lm},
\end{equation}
Eq.(\ref{radial}) can be rewritten in the form of a Schr\"{o}dinger-like wave equation
\begin{equation}
	\frac{\mathrm{d}^{2} \psi }{\mathrm{d} r^{2} } +\left [ (1+L)\omega ^{2}-V  \right ] =0,
\end{equation}
where
\begin{equation}
	(1+L)\omega^{2}-V=\frac{U(1+L)^2 +M^2 -\tilde{a}^2}{\tilde{\Delta}^2 (1+L)^2}.
\end{equation}
For large $r$, the effective potential and its derivative with respect to $r$ are
\begin{equation}
	V=\mu^2 (1+L)-\frac{2(1+L)M(2\omega^2 -\mu^2)}{r}+\mathcal{O}(\frac{1}{r^2})
\end{equation}
and
\begin{equation}
	V'=\frac{2(1+L)M(2\omega^2 -\mu^2)}{r^2}+\mathcal{O}(\frac{1}{r^3}),
\end{equation}
respectively. A trapping well exists if $V'\to 0^+$ as $r\to\infty$, which means that
\begin{equation}\label{3}
	2\omega^2 -\mu^2 >0.
\end{equation}
Therefore, the frequency regime for the bound states is independent of LSB parameter $L$ and is given by
\begin{equation}\label{bound}
	\frac{\mu }{\sqrt{2}} < \omega < \mu .
\end{equation}

It is worth noting that in Ref.\cite{Khodadi:2021owg} the author obtained a different result about the frequency regime due to an equation decomposition different from Eq.\eqref{radial} of this work. By careful checking, we make sure our equation decomposition (Eqs.\eqref{radial}-\eqref{angular}) is in accordance with the results in mathematical handbooks and can also be confirmed by the recent work about the quasinormal modes of this Kerr-like black hole\cite{Kanzi:2021cbg}. We argue that the decomposition of the field equation (Eq.(3.3)) in Ref.\cite{Khodadi:2021owg} is erroneous in that it apparently lacks the highly nontrivial angular eigenvalues $A_{lm}$. Consequently, the incorrect equation decomposition in Ref.\cite{Khodadi:2021owg} leads to an incorrect asymptotic behavior of the radial function, which is inconsistent with Eq.\eqref{boundary condition} in this work and with other related work\cite{Ding_2021}.

\section{Bound state spectrum}\label{Sec.4}
For a given parameter $L$ and a scalar field with mass $\mu$, the boundary condition\eqref{boundary condition} singles out a
discrete set of bound state frequencies $\left\lbrace \omega_{n}\right\rbrace$. In this section, we compute the bound state frequencies using the continued-fraction method. To this end, the asymptotic behavior of the radial function is written as
\begin{equation}\label{4}
	R_{lm}(r\to r_{+})\sim(r-r_{+})^{-i\sigma}
\end{equation}
and
\begin{equation}\label{5}
	R_{lm}(r\to\infty)\sim r^{\chi-1}e^{kr},
\end{equation}
where
\begin{equation}
	\sigma =\frac{r_{+}^{2}+\tilde{a}^{2}}{r_{+}-r_{-}} (\omega -\omega _{c} )\sqrt{1+L},
\end{equation}
\begin{equation}
	\chi =\frac{M(\mu ^{2}-2\omega ^{2})(1+L)}{k} ,
\end{equation}
and
\begin{equation}
	k=-\sqrt{(\mu^{2}-\omega^{2})(1+L)}.
\end{equation}
The solution of radial equation can be expressed as
\begin{equation}\label{2}
	R_{lm}=(r-r_{+})^{-i\sigma}(r-r_{-})^{\chi-1+i\sigma}e^{kr}\sum_{n=0}^{\infty}a_{n}(\frac{r-r_{+}}{r-r_{-}})^{n}
\end{equation}
so that both Eqs.\eqref{4} and\eqref{5} can be accommodated. Substituting (\ref{2}) into Eq.(\ref{radial}), one obtains a three-term recurrence relation for the coefficients $a_{n}$,
	\begin{align}
	&\alpha _{0}a_{1}+\beta _{0}a_{0}=0\label{a1a0},\\
	&\alpha _{n}a_{n+1}+\beta _{n}a_{n}+\gamma _{n}a_{n-1}=0  ,\space n=1,2,3\cdots  ,
	\end{align}
where
\begin{equation}
	\begin{aligned}
		&\alpha_{n}=n^{2}+(c_{0}+1)n+c_{0},\\
		&\beta_{n}=-2n^{2}+c_{1}+c_{2},\\
		&\gamma_{n}=n^{2}+c_{3}n+c_{4}.
	\end{aligned}
\end{equation}
The constants $c_{0},c_{1},c_{2},c_{3}$ and $c_{4}$ are given by
\begin{equation}\label{40}
	\begin{aligned}
	c_{0}=&1-2Mi\omega \sqrt{1+L}-\frac{i}{\eta} \left(2M^{2} \omega \sqrt{1+L}-\tilde{a}m\sqrt{1+L} \right),\\
	c_{1}=&-2c_{0}+2(M+2\eta)k-\frac{2M\omega ^{2}(1+L)}{k},\\
	c_{2}=&-A_{lm}(1+L)-(M+\eta)^{2}k^{2}+2(M+\eta)M\omega ^{2}(1+L)\\
	&-2Mi\omega \sqrt{1+L}-c_{0}\bigg[ -(M+\eta)k+\frac{M\omega ^{2}(1+L)}{k}\\
	&+1-2Mi\omega \sqrt{1+L}\bigg],\\
	c_{3}=&-2Mi\omega \sqrt{1+L}-\frac{i}{\eta}\bigg(2M^{2} \omega\sqrt{1+L}-\tilde{a}m\sqrt{1+L}\bigg)\\
	&-\frac{2M(\mu ^{2}-2\omega^{2})(1+L)}{k},\\
	c_{4}=&(k+i\omega\sqrt{1+L})^{2}M\bigg[ \frac{i}{\eta} \bigg(2M^{2}\omega \sqrt{1+L}-\tilde{a}m\sqrt{1+L} \bigg)\\
	&+\frac{M(\mu ^{2}-2\omega ^{2})(1+L)}{k} \bigg]/k	,
	\end{aligned}
\end{equation}
where
\begin{equation}
	\eta=\sqrt{M^{2}-\tilde{a}^{2}}.
\end{equation}
One can see that the factor $(1+L)$ or $\sqrt{1+L}$ appears many times in Eq.\eqref{40} and cannot be absorbed in $\tilde{a}$. This again roots in the fact that $\tilde{a}$ and $\tilde{\Delta}$ given in Eq.\eqref{metric3} have a different relation from their counterparts in the Kerr case and $\tilde{\Delta}$ involves explicitly the LSB parameter $L$. At last, the radial equation\eqref{radial} involving $\tilde{\Delta}$ results in the multiple presence of $L$ in Eq.\eqref{40}, which shows that $L$ will play a significant role in the bound state spectrum. It is easy to check that when $L=0$, the expressions for $\alpha_{n},\beta_{n}$ and $\gamma_{n}$ reduce to Kerr case given in Ref.\cite{Dolan_2007}.

The ratio of successive coefficients $a_{n}$ is given by an infinite generalized continued fraction
\begin{equation}
	\frac{a_{n+1}}{a_{n}}=-\frac{\gamma_{n+1}}{\beta_{n+1}-}\frac{\alpha_{n+1}\gamma_{n+2}}{\beta_{n+2}-}\frac{\alpha_{n+2}\gamma_{n+3}}{\beta_{n+3}-}\cdots .
\end{equation}
Substituting $n=0$ into the above expression and comparing with Eq.(\ref{a1a0}), one can obtain the characteristic equation
\begin{equation}\label{characteristic equation}
	\beta _{0}-\frac{\alpha _{0}\gamma_{1}}{\beta_{1}-}\frac{\alpha_{1}\gamma_{2}}{\beta_{2}-}\frac{\alpha_{2}\gamma_{3}}{\beta_{3}-}\dots=0.
\end{equation}
This equation is only satisfied for particular values of $\omega$ corresponding to the frequencies of the bound state modes. We compute these frequencies numerically by minimizing the absolute value of the left side of Eq.\eqref{characteristic equation}. For a given bound state mode, the resulting frequency $\omega$ is generally a complex number. The real part of it $(\omega_{R})$ represents the oscillating frequency of the field, while the imaginary part $(\omega_{I})$ indicates the evolution of the wave amplitude. $\omega_{I}$ being negative or positive means a decaying or growing scalar field, respectively. And the magnitude of |$\omega_{I}$| represents the corresponding damping or growth rate. Among the growing modes of the field, if exist, the ones with $l=m=1$ grow most rapidly. Therefore, we will focus on the behavior of these modes in the following discussion.

\begin{figure}
	\centering
	\includegraphics[width=10cm,height=10cm]{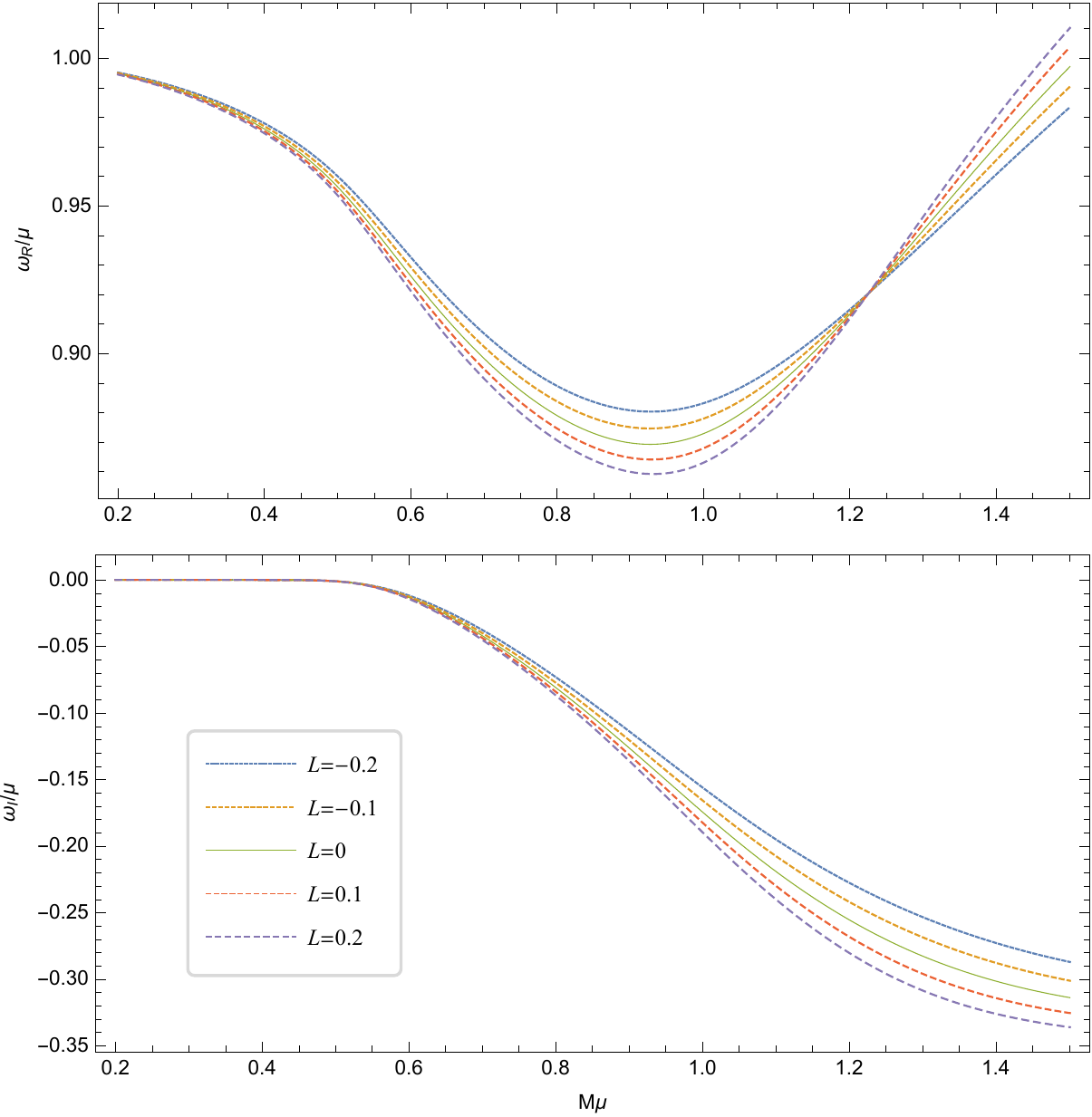}
	\caption{Bound state frequencies versus the mass coupling $M\mu$ for different parameter $L$, where we have set $l=m=1$ and $\tilde{a}/M=0.9$.}\label{Fig.1}
\end{figure}

Fig.\ref{Fig.1} shows how the bound state frequency of the state with $n=0,l=m=1$ and $\tilde{a}/M=0.9$ vary with the mass coupling $M\mu$ and LSB parameter $L$. The upper panel illustrates the real part of the frequency $(\omega_{R})$. One immediately observes that $\omega_{R}/\mu$ has a minimum at some $M\mu$, which corresponds to the maximum binding energy. It is obvious that the maximum binding energy increases with $L$. For example, when $L=0$ (Kerr case), the maximum binding energy is around $13.1\%$ of the rest mass energy. It becomes $\sim 11.97\%$ for $L=-0.2$ and $\sim 14.08\%$ for $L=0.2$. The lower panel illustrates the imaginary part of the frequency $(\omega_{I})$. It shows that the damping rate increases with $L$. Moreover, $\omega_{I}$ is actually positive at low couplings, which we cannot discern in this figure but is just the regime of the superradiant instability we will analyze in the following section.

\section{Superradiant instability}
In this section, we focus on unstable modes in which the imaginary part of the frequency is positive ($\omega_{I}>0$). This happens when Eqs.\eqref{1} and \eqref{bound} are both satisfied.

\begin{figure}[htbp]
	\centering
	\includegraphics[width=10cm,height=10cm]{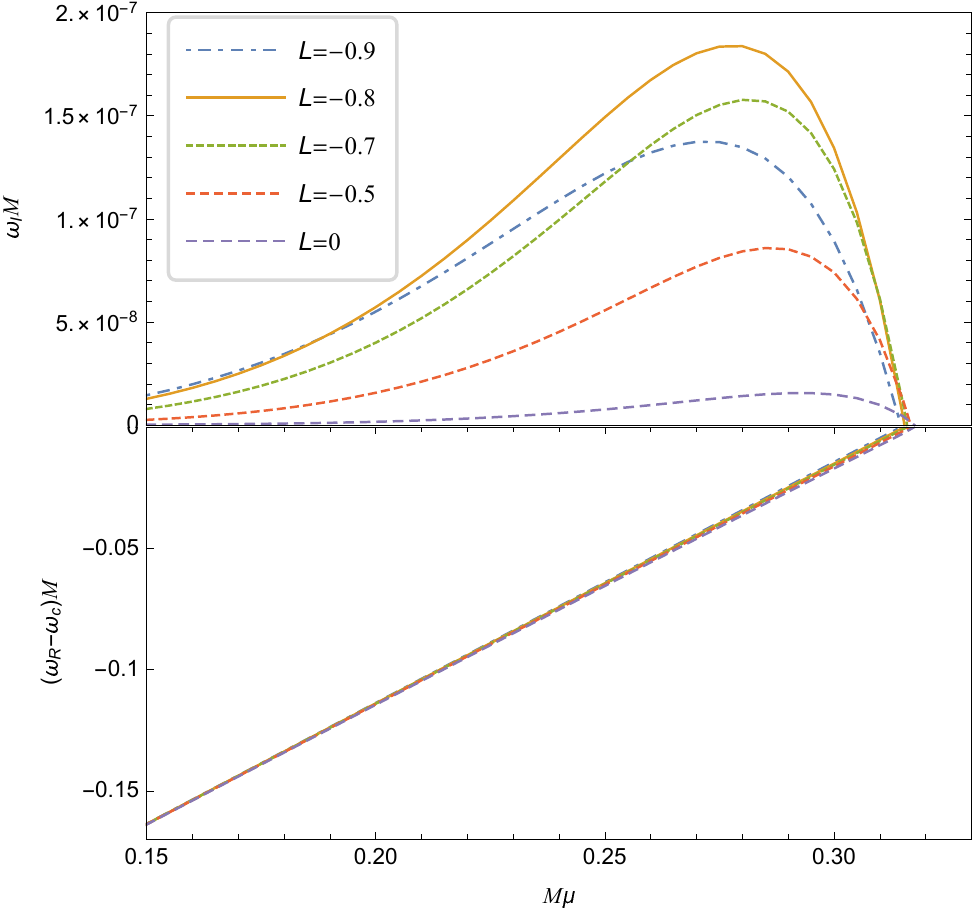}
	\caption{Bound state spectrum of a scalar field  ($l=m=1$) for different values of $L$ when $\tilde{a}/M=0.9$.}\label{Fig.2}
\end{figure}

\begin{figure}[htbp]
	\centering
	\includegraphics[width=10cm,height=10cm]{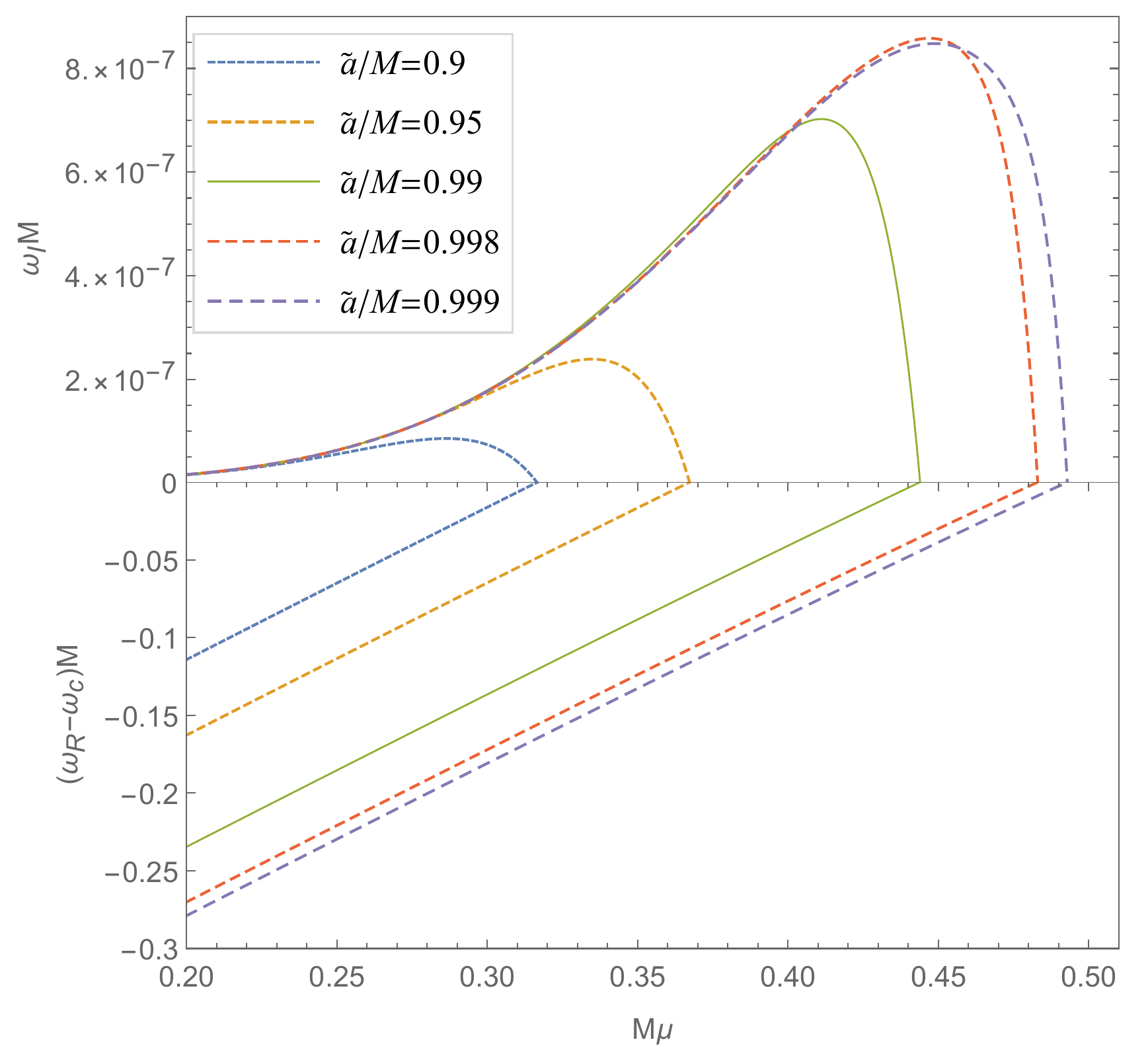}
	\caption{Bound state spectrum of a scalar field  ($l=m=1$) for different values of $\tilde{a}$  when $L=-0.5$.}\label{Fig.3}
\end{figure}

Fig.\ref{Fig.2} and Fig.\ref{Fig.3} show the detailed behaviors of the positive $\omega_{I}$ versus the mass coupling $M\mu$ for different $L$ and $\tilde{a}$. In Fig.\ref{Fig.2}, we take $\tilde{a}/M=0.9$ and the curves correspond to different $L$. And in Fig.\ref{Fig.3}, we take $L=-0.5$ and the curves correspond to different $\tilde{a}$. We also plot the $\omega_{R}$ compared to the critical frequency $\omega_{c}$ of superradiance in this regime. Recall that superradiance occurs if and only if $\omega<\omega_{c}$. It is obvious from the two plots that $\omega_{I}$ is positive when $\omega_{R}<\omega_{c}$. It is then confirmed that the growth
of the field comes from the superradiance mechanism.

\begin{table}[htbp]\label{table1}
	\centering
	\caption{Maximum instability growth rate of different $L$ with $l=m=1$.}
	\begin{tabular}{|c|l|l|l|}
		\hline
		$L$&\multicolumn{1}{|c|}{$\tilde{a}^{*}/M$}&\multicolumn{1}{|c|}{$M\mu^{*}$}&\multicolumn{1}{|c|}{$\omega_{I}^{*}M$}\\
		\hline
		-0.95&0.99962&0.425919&5.28603$\times10^{-7}$\\
		\hline
		-0.9&0.99932&0.432026&1.20933$\times10^{-6}$\\
		\hline
		-0.8&0.99886&0.439019&1.67536$\times10^{-6}$\\
		\hline
		-0.7&0.99848&0.442775&1.48849$\times10^{-6}$\\
		\hline
		-0.5&0.99786&0.446469&8.58337$\times10^{-7}$\\
		\hline
		-0.4&0.99759&0.447413&6.23684$\times10^{-7}$\\
		\hline
		-0.3&0.99733&0.448040&4.50565$\times10^{-7}$\\
		\hline
		-0.2&0.99708&0.448460&3.25704$\times10^{-7}$\\
		\hline
		-0.1&0.99685&0.448736&2.36348$\times10^{-7}$\\
		\hline
		0&0.99663&0.448901&1.72440$\times10^{-7}$\\
		\hline
		0.1&0.99642&0.448995&1.26593$\times10^{-7}$\\
		\hline
		0.2&0.99622&0.449046&9.35381$\times10^{-8}$\\
		\hline
		0.3&0.99603&0.449049&6.95651$\times10^{-8}$\\
		\hline
		0.4&0.99584&0.449008&5.20677$\times10^{-8}$\\
		\hline
		0.5&0.99567&0.448966&3.92135$\times10^{-8}$\\
		\hline
	\end{tabular}
\end{table}

One can also see from the two plots that each curve of $\omega_{I}$ has a peak and there is a highest peak of $\omega_{I}$ corresponding to $L=-0.8$ in Fig.\ref{Fig.2} and $\tilde{a}/M=0.998$ in Fig.\ref{Fig.3}. That is, for a given $\tilde{a}$ or $L$, there may be a certain set of parameters that leads to a maximum growth rate $\omega_{I}^{*}$. Then, taking different $L$, we search for the $\omega_{I}^{*}$, as well as the corresponding set of parameters $\tilde{a}^{*}$ and $M\mu^{*}$ by Monte Carlo method. The results are shown in Table \uppercase\expandafter{\romannumeral1} where the parameters with an asterisk represent their values corresponding to the maximum growth rate. It is clear that the result for Kerr black hole, the maximum growth rate $\omega_{I}^{*}M=1.72440\times10^{-7}$ at $a^{*}/M=0.99663$ and $M\mu^{*}=0.448901$\cite{Dolan_2013}, is recovered when $L=0$. One can see that as $L$ increases, the rotation speed $\tilde{a}^{*}$ required to reach the maximum growth rate decreases, while the mass coupling $M\mu^{*}$ increases first and then decreases although it does not change significantly. Most importantly, the maximum growth rate $\omega_{I}^{*}$ does not vary monotonously. It is suspected that there may exist an overall maximum growth rate when all three of the parameters of $L$, $\tilde{a}$ and $M\mu$ are taken into account, which corresponds to the most unstable mode of the field. Using the Monte Carlo method, we find the most unstable mode occurs at $L=-0.79637$, $\tilde{a}/M = 0.99884$ and $M\mu=0.43920$, with $\omega_{I}M=1.676\times10^{-6}$. It is about 10 times of the maximum growth rate in Kerr black hole.

\section{Conclusion and discussions}
In this paper, we have looked into the instability of the Kerr-like black hole in the Einstein-bumblebee gravity by considering the perturbation of a massive scalar field. Using a suitable redefinition of the spin parameter $\tilde{a}=\sqrt{1+L}a$, we rewrite the metric into a form most resembling to the Kerr solution in GR. In this form,  $\tilde{a}$ plays a similar role to the Boyer-Lindquist parameter $a$ in Kerr metric. The effect of the LSB parameter $L$ via the physical quantities such as the horizon radii or angular velocity cannot be separated from $\tilde{a}$ and hence is not trackable. It follows that the frequency regimes of superradiance and bound states are unaffected by $L$. However, the effect of $L$ on the metric cannot be fully erased by redefining $\tilde{a}$ in that $L$ still appears explicitly in $\tilde{\Delta}$. Our analysis shows that the bound state spectrum and superradiance are indeed affected by $L$. In particular, we calculate the bound state spectrum via the continued-fraction method and show the influence of LSB parameter $L$ on the maximum binding energy and the damping rate. The superradiant instability could occur for this black hole since the superradiance condition and the bound state condition could be both satisfied. We find the growth rate of the field does not depend monotonously on $L$. In view of a parameter space spanned by the LSB parameter $L$, the rotation parameter $\tilde{a}$ and the mass coupling $M\mu$, there exists an overall maximum growth rate $\omega_{I}$, which corresponds to the most unstable mode of the field. By Monte Carlo method, we have found that the most unstable mode occurs at $L=-0.79637$, $\tilde{a}/M= 0.99884$ and $M\mu=0.43920$ for $l=m=1$ state, with $\omega_{I}M=1.676\times10^{-6}$. It is about 10 times of that in Kerr black hole.

Recent studies show that when nonlinear effects are taken into account, the instability of Kerr black hole will result in a rotating black hole embedded in a massive bosonic field that orbits around the black hole\cite{Herdeiro_2014,Herdeiro_2016,East_2017}. Therefore, it is expected that the superradiant instability discussed in the present work may lead to a Kerr-like black hole with scalar hair, which is a further work we intend to study in the future.

\section*{Acknowledgement}\label{ackn}
We thank Dr. Yang Huang for useful suggestions.
This work is supported by the National 
Science Foundation of China under Grant No. 12105179.

Rui Jiang and Rui-Hui Lin contribute equally to this work.
	\bibliography{ref}
\end{document}